\documentclass[1p,sort&compress]{elsarticle}

\usepackage{graphicx}
\usepackage{subfig}
\usepackage{physics}
\usepackage{tikz}
\usetikzlibrary{arrows,shapes,trees,backgrounds,snakes,calc,arrows.meta} 
\usepackage{siunitx}

\usepackage{lineno,hyperref}
\modulolinenumbers[5]

\journal{Journal of \LaTeX\ Templates}

\hypersetup{
  colorlinks = true,
}

\begin{document}
\begin{frontmatter}

\title{Feasibility study of the time reversal symmetry tests in decays of metastable positronium atoms with the J-PET detector}

\cortext[cor1]{Corresponding author.}
\author[WFAIS]{A.~Gajos}
\author[INFN]{C.~Curceanu}
\author[WFAIS]{E.~Czerwi\'nski}
\author[WFAIS]{K.~Dulski}
\author[LUBLIN]{M.~Gorgol}
\author[WFAIS]{N.~Gupta-Sharma}
\author[VIENNA]{B.C.~Hiesmayr}
\author[LUBLIN]{B.~Jasi\'nska}
\author[WFAIS]{K.~Kacprzak}
\author[WFAIS,PAN]{\L.~Kap\l on}
\author[WFAIS]{D.~Kisielewska}
\author[WFAIS]{G.~Korcyl} 
\author[SWIERK]{P.~Kowalski}
\author[WFAIS]{T.~Kozik}
\author[NCBJ]{W.~Krzemie\'n}
\author[WFAIS]{E.~Kubicz}
\author[WFAIS,MOSUL]{M.~Mohammed}
\author[WFAIS]{Sz.~Nied\'zwiecki}
\author[WFAIS]{M.~Pa\l ka}
\author[WFAIS]{M.~Pawlik-Nied\'zwiecka}
\author[SWIERK]{L.~Raczy\'nski}
\author[WFAIS]{J.~Raj}
\author[WFAIS]{Z.~Rudy}
\author[WFAIS]{S.~Sharma}
\author[WFAIS]{Shivani}
\author[SWIERK]{R.~Shopa}
\author[WFAIS]{M.~Silarski}
\author[WFAIS]{M.~Skurzok}
\author[SWIERK]{W.~Wi\'slicki}
\author[LUBLIN]{B.~Zgardzi\'nska}
\author[WFAIS]{M.~Zieli\'nski}
\author[WFAIS]{P.~Moskal\corref{cor1}}
\ead{p.moskal@uj.edu.pl}
\address[WFAIS]{Faculty of Physics, Astronomy and Applied Computer Science,
 Jagiellonian University, 30-348 Cracow, Poland}
\address[INFN]{INFN, Laboratori Nazionali di Frascati, 00044, Frascati, Italy}
\address[LUBLIN]{Department of Nuclear Methods, Institute of Physics, Maria Curie-Sk\l{}odowska University, 20-031, Lublin, Poland.}
\address[VIENNA]{Faculty of Physics, University of Vienna, 1090, Vienna, Austria}
\address[PAN]{Institute of Metallurgy and Materials Science of Polish Academy of Sciences, Cracow, Poland.}
\address[SWIERK]{\'Swierk Department of Complex Systems, National Centre for Nuclear Research, 05-400 Otwock-\'Swierk, Poland}
\address[NCBJ]{High Energy Department, National Centre for Nuclear Research, 05-400 Otwock-\'Swierk, Poland}
\address[MOSUL]{Department of Physics, College of Education for Pure Sciences, University of Mosul, Mosul, Iraq}

\begin{abstract}
  This article reports on the feasibility of testing of the symmetry under reversal in time in a purely leptonic system constituted by positronium atoms using the J-PET detector. The present state of $\mathcal{T}$ symmetry tests is discussed with an emphasis on the scarcely explored sector of leptonic systems. Two possible strategies of searching for manifestations of $\mathcal{T}$ violation in non-vanishing angular correlations of final state observables in the decays of metastable triplet states of positronium available with J-PET are proposed and discussed.
  Results of a pilot measurement with J-PET and assessment of its performance in reconstruction of three-photon decays are shown along with an analysis of its impact on the sensitivity of the detector for the determination of $\mathcal{T}$-violation sensitive observables.
\end{abstract}

\end{frontmatter}


\section*{Introduction}
The concept of symmetry of Nature under discrete transformations has been exposed to numerous experimental tests ever since its introduction by E.~Wigner in 1931~\cite{wigner1931}.
The first evidence of violation of the supposed symmetries under spatial~($\mathcal{P}$) and charge~($\mathcal{C}$) parity transformations in the weak interactions has been found already in 1956 and 1958 respectively~\cite{parity_violation, c_violation}. However, observation of noninvariance of a physical system under reversal in time required over 50 years more and was finally performed in the system of entangled neutral B mesons in 2012~\cite{t_violation_babar}.
Although many experiments proved violation of the combined $\mathcal{CP}$ symmetry, leading to $\mathcal{T}$ violation expected on the ground of the $\mathcal{CPT}$ theorem, experimental evidence for noninvariance under time reversal remains scarce to date.

The Jagiellonian PET (J-PET) experiment aims at performing a test of the symmetry under reversal in time in a purely leptonic system constituted by ortho-positronium (\mbox{o-Ps}) with a precision unprecedented in this sector. The increased sensitivity of J-PET with respect to previous discrete symmetry tests with \mbox{o-Ps$\to$3$\gamma$} is achieved by a large geometrical acceptance and angular resolution of the detector as well as by improved control of the positronium atoms polarization. In this work, we report on the results of feasibility studies for the planned $\mathcal{T}$ violation searches by determination of angular correlations in the o-Ps$\to 3\gamma$ decays based on a test run of the J-PET detector.

This article is structured as follows:
next section briefly discusses the properties of time and time reversal in quantum systems. Subsequently,
Section~\ref{sec:status} provides an overview of the present status and available techniques of testing of the symmetry under reversal in time and points out the goals of the J-PET experiment in this field.
A brief description of the detector and details of the setup used for a test measurement are given in Section~\ref{sec:jpet}. Section~\ref{sec:strategies} discusses possible strategies to test the time reversal symmetry with J-PET.\@
Results of the feasibility studies are presented in Section~\ref{sec:test} and their impact on the perspectives for a $\mathcal{T}$ test with J-PET is discussed in Section~\ref{sec:summary}.

\section{Time and reversal of physical systems in time}\label{sec:time}
Although the advent of special relativity made it common equate time with spatial coordinates, time remains a distinct concept. Its treatment as an external parameter used in classical mechanics still cannot be consistently avoided in today's quantum theories~\cite{muga}. As opposed to position and momentum, time lacks a corresponding operator in standard quantum mechanics and thus, countering the intuition, cannot be an observable. Moreover, a careful insight into the time evolution of unstable quantum systems reveals a number of surprising phenomena such as deviations from exponential decay law~\cite{urbanowski_decay, giacosa_unstable} or emission of electromagnetic radiation at late times~\cite{Urbanowski:2013tfa}. The decay process, inevitably involved in measurements of unstable systems is also a factor restricting possible studies of the symmetry under time reversal~\cite{bernabeu_colloquium}.

While efforts are taken to define a time operator, observation of $\mathcal{CP}$ violation in the decaying meson systems disproves certain approaches~\cite{hiesmayr_didomenico}. Alternatively, concepts of time intervals not defined through an external parameter may be considered using tunneling and dwell times~\cite{kelkar_quantum_time, kelkar_dwell_times}. However, also in this case invariance under time reversal is an important factor~\cite{Kelkar:2008na}. 

It is important to stress that all considerations made herein are only valid if gravitational effects are not considered. In the framework of general relativity with a generic curved spacetime, the concept of inversion of time (as well as the $\mathcal{P}$ transformation) loses its interpretation specific only to the linear affine structure of spacetime~\cite{Sokolowski:2017yoc}.

The peculiar properties of time extend as well to the operation of reversing physical systems in time (the T operator), which results in grave experimental challenges limiting the possibilities of $\mathcal{T}$ violation measurements. In contrast to the unitary P and C operators, T can be shown to be antiunitary. As a consequence, no conserved quantities may be attributed to the T operation~\cite{sachs} excluding symmetry tests by means of e.g.\ testing selection rules.

Feasibility of $\mathcal{T}$ tests based on a comparison between time evolution of a physical system in two directions, i.e. $\ket{\psi(t)}\to \ket{\psi(t+\delta t)}$ and $\ket{\psi(t+\delta t)}\to \ket{\psi(t)}$ is also limited as most of the processes which could be used involve a decaying state making it impractical to obtain a reverse process with the same conditions in an experiment. The only exception exploited to date is constituted by transitions of neutral mesons between their flavour-definite states and CP eigenstates~\cite{babar_theory, theory:bernabeu-t}. A comparison of such reversible transitions in a neutral B meson system with quantum entanglement of $\mathrm{B}^0\overline{\mathrm{B^0}}$ pairs produced in a decay of $\Upsilon(4s)$ yielded the only direct experimental evidence of violation of the symmetry under reversal in time obtained to date~\cite{t_violation_babar}. While a similar concept of $\mathcal{T}$~violation searches is currently pursued with the neutral kaon system~\cite{theory:bernabeu-t,Gajos:2015ija, Gajos:2017tlb}, no direct tests  of this symmetry have been proposed outside the systems of neutral mesons.

%
%
In the absence of conserved quantities and with the difficulties of comparing mutually reverse time evolution processes in decaying systems, manifestations of  $\mathcal{T}$ violation may still be sought in non-vanishing expectation values of certain operators odd under the T transformation~\cite{wolfenstein_summary}. It follows from the antiunitarity of the T operator that for any operator $\mathcal{O}$:
\begin{equation}
  \mel{\phi}{\mathcal{O}}{\psi} = \mel{\phi}{T^{\dagger}T\mathcal{O}T^{\dagger}T}{\psi}  = \mel{\phi_T}{\mathcal{O}_T}{\psi_T}^*,
\label{eq:rel_any_operator}
\end{equation}
where the $T$ subscript denotes states and operators transformed by the operator of reversal in time.
Therefore, an operator odd with respect to the T transformation (i.e. $\mathcal{O}_T = -\mathcal{O}$) must satisfy:
\begin{equation}
  \mel{\phi}{O}{\psi} = {-}\mel{\phi_T}{O}{\psi_T}^*.
\end{equation}
For stationary states, or in systems where conditions on interaction dynamics such as absence of significant final state interactions are satisfied~\cite{PhysRevC.52.1041}, the mean value of a  a T-odd and hermitian operator must therefore vanish in case of $\mathcal{T}$~invariance:
\begin{equation}
  \expval{O}_T = {-}\expval{O},
  \label{eq:todd}
\end{equation}
and violation of the $\mathcal{T}$ symmetry may thus be manifested as a non-zero expectation value of such an operator.

\section{Status and strategies of $\mathcal{T}$ symmetry testing}\label{sec:status}
A number of experiments based on the property of T operator demonstrated in Equations~\ref{eq:rel_any_operator}-\ref{eq:todd} have been conducted to date. The electric dipole moment of elementary systems, constituting a convenient T-odd operator, has been sought for neutrons and electrons in experiments reaching a precision of $10^{-26}$ and $10^{-28}$ respectively~\cite{nedm,eedm}. However, none of such experiments has observed $\mathcal{T}$ violation to date despite their excellent sensitivity. In another class of experiments, a T-odd operator is constructed out of final state observables in a decay process, such as the weak decay \mbox{$K^+\to\pi^0\mu^+\nu$} studied by the KEK-E246 experiment~\cite{operatory_kek} in which the muon polarization transverse to the decay plane ($\mathcal{P}_T = \mathcal{P}_{K}\cdot (\mathbf{p}_{\pi}\times\mathbf{p}_{\mu}) /|\mathbf{p}_{\pi}\times\mathbf{p}_{\mu}|$) was determined as an observable whose non-zero mean value would manifest $\mathcal{T}$ violation. However, neither this measurement nor similar studies using decays of polarized $^8$Li nuclei~\cite{bodek_li} and of free neutrons~\cite{bodek_free_n} have observed significant mean values of T-odd final state observables.

Notably, although the property of reversal in time shown in Equations~\ref{eq:rel_any_operator}-\ref{eq:todd} is not limited to any particular system nor interaction, is has been mostly exploited to test the $\mathcal{T}$ symmetry in weak interactions. Whereas the latter is the most promising candidate due to well proven $\mathcal{CP}$ violation, evidence for $\mathcal{T}$ noninvariance may be sought in other physical systems and phenomena using the same scheme of a symmetry test.
Systems constituted by purely leptonic matter are an example of a sector where experimental results related to the time reversal symmetry --- and to discrete symmetries in general --- remain rare.
Several measurements of neutrino oscillations are being conducted by the NO$\nu$A and T2K experiments searching for $\mathcal{CP}$ violation in the $\nu_{\mu}\to \nu_{e}$ and $\bar{\nu}_{\mu}\to\bar{\nu}_{e}$ channels~\cite{nova, neutrino_oscillations}, which may provide indirect information on the $\mathcal{T}$ symmetry. Another notable test of discrete symmetries in the leptonic sector is the search for the violation of Lorentz and $\mathcal{CPT}$ invariance based on the Standard Model Extension framework~\cite{Kostelecky:2001ff} and anti-CPT theorem~\cite{Greenberg:2002uu} which has also been performed by T2K~\cite{neutrino_lorentz, Kostelecky:2008ts}.
Other possible tests of these symmetries with the positronium system include spectroscopy of the 1S-2S transition~\cite{PhysRevA.60.2792} and measuring the free fall acceleration of positronium~\cite{Kostelecky:2015nma}.
However, the question of the  $\mathcal{T}$, $\mathcal{CP}$ and $\mathcal{CPT}$ symmetries in the leptonic systems remains open as the aforementioned experiments have not observed a significant signal of a violation.

Few systems exist which allow for discrete symmetry tests in a purely leptonic sector. However, a candidate competitive with respect to neutrino oscillations is constituted by the electromagnetic decays of positronium atoms, exotic bound states of an electron and a positron. With a reduced mass only twice smaller than that of a hydrogen atom, positronium is characterized by a similar energy level structure. At the same time, it is a metastable state with a lifetime strongly dependent on the spin configuration. The singlet state referred to as para-positronium, may only decay into an even number of photons due to charge parity conservation, and has a lifetime (in vacuum) of $0.125$~ns. The triplet state (ortho-positronium, o-Ps) is limited to decay into an odd number of photons and lives in vacuum over three orders of magnitude longer than the singlet state ($\tau_{o-Ps}=142$~ns)~\cite{PhysRevLett.72.1632,PhysRevLett.90.203402,JINNOUCHI2003117}.

Being an eigenstate of the parity operator alike atoms, positronium is also characterized by symmetry under charge conjugation typical for particle\discretionary{-}{-}{-}antiparticle systems. Positronium atoms are thus a useful system for discrete symmetry studies. Moreover, they may be copiously produced in laboratory conditions using typical sources of $\beta^+$ radiation~\cite{Jasinska:2016qsf}, giving positronium-based experiments a technical advantage over those using e.g. aforementioned neutrino oscillations. However, few results on the discrete symmetries in the positronium system have been reported to date. The most precise measurements studied the angular correlation operators in the decays of ortho-positronium states into three photons and determined mean values of final state operators odd under the $\mathcal{CP}$ and $\mathcal{CPT}$ conjugations, finding no violation signal at the sensitivity level of $10^{-3}$~\cite{cp_positronium,cpt_positronium}.
Although the aforementioned studies sought for violation of $\mathcal{CP}$ and $\mathcal{CPT}$, it should be emphasized that the operators used therein were odd under the T~operation as well, leading to an implicit probe also for the symmetry under reversal in time.


The results obtained to date, showing no sign of violation, were limited in precision by technical factors such as
detector geometrical acceptance and resolution, uncertainty of positronium polarization and data sample size. In terms of physical restrictions, sensitivity of such discrete symmetry tests with ortho-positronium decays is only limited by possible false asymmetries arising from photon-photon final state interactions at the precision level of $10^{-9}$~\cite{Arbic:1988pv,Bernreuther:1988tt}. The J-PET experiment thus sets its goal to explore the $\mathcal{T}$-violating observables at precision beyond the presently established $10^{-3}$ level~\cite{moskal_potential}.

\section{The J-PET detector}\label{sec:jpet}
The J-PET (Jagiellonian Positron Emission Tomograph) is a photon detector constructed entirely with plastic scintillators. Along with constituting the first prototype of plastic scintillator-based cost-effective PET scanner with a large field of view~\cite{moskal_patent, Niedzwiecki:2017nka}, it may be used to detect photons in the sub-MeV range such as products of annihilation of positronium atoms, thus allowing for a range of studies related to discrete symmetries and quantum entanglement~\cite{moskal_potential}.

\begin{figure}[h!]
  \centering
  \includegraphics[width=0.45\textwidth]{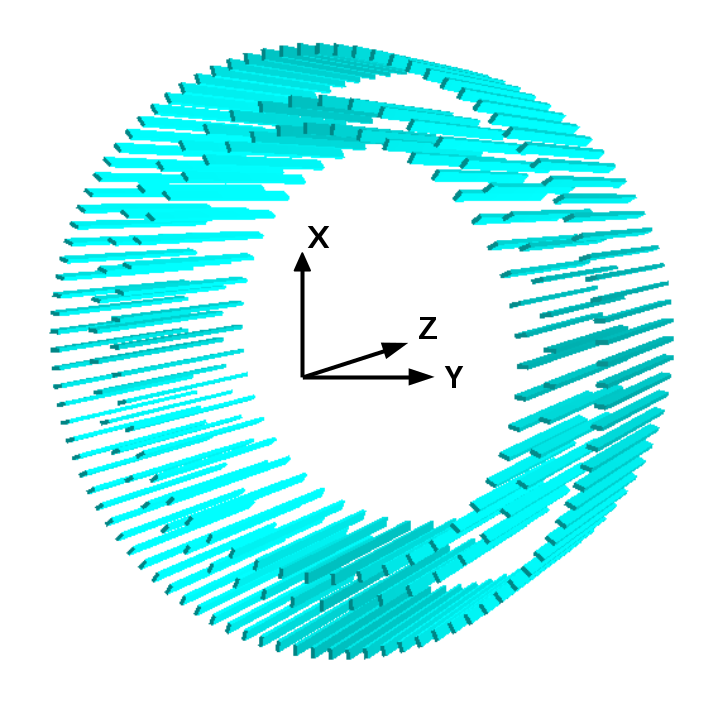}
  \caption{Schematic view of the J-PET detector consisting of 192 plastic scintillator strips arranged in three concentric layers with radii ranging from 42.5 cm to 57.5 cm. The strips are oriented along the Z axis of the detector barrel.}\label{fig:jpet}
\end{figure}

J-PET consists of three concentric cylindrical layers of axially arranged $\gamma$~detection modules based on strips of EJ-230 plastic scintillator. Each scintillator strip is 50~cm long with a rectangular cross-section of 7$\times$19~mm$^2$. Within a detection module, both ends of a scintillator strip are optically coupled to photomultiplier tubes. Due to low atomic number of the elements constituting plastic scintillators, $\gamma$ quanta interact mostly through Compton scattering in the strips, depositing a part of their energy dependent on the scattering angle. The lack of exact photon energy determination in J-PET is compensated by fast decay time of plastic scintillators resulting in high time resolution and allowing for use of radioactive sources with activity as high as 10~MBq. The energy deposited by photons scattered in a scintillator is converted to optical photons which travel to both ends of a strip undergoing multiple internal reflections. Consequently, the position of $\gamma$~interaction along a detection module is determined using the difference between effective light propagation times to the two photomultiplier tubes attached to a scintillator strip~\cite{jpet_single_module}. In the transverse plane of the detector, $\gamma$ interactions are localized up to the position of a single module, resulting in an azimuthal angle resolution of about 1$^{\circ}$.

Although the J-PET $\gamma$ detection modules do not allow for a direct measurement of total photon energy, recording interactions of all photons from a $3\gamma$~annihilation allows for an indirect reconstruction of photons' momenta based on event geometry and 4-momentum conservation~\cite{daria_epjc}.

As J-PET is intended for a broad range of studies from medical imaging~\cite{monika_2gamma_imaging} through quantum entanglement~\cite{Hiesmayr:2017xgx,Nowakowski:2017jbr} to tests of discrete symmetries~\cite{moskal_potential}, its data acquisition is operating in a triggerless mode~\cite{greg_daq} in order to avoid any bias is the recorded sample of events. Electric signals produced by the photomultipliers are sampled in the time domain at four predefined voltage thresholds allowing for an estimation of the deposited energy using the time over threshold technique~\cite{marek_fee}. Further reconstruction of photon interactions as well as data preselection and handling is performed with dedicated software~\cite{Krzemien:2015hkb,Krzemien:2015xea}. Several extensions of the detector are presently in preparation such as improvements of the J-PET geometrical acceptance by inclusion of additional detector layers~\cite{daria_epjc} as well as enhanced scintillator readout with silicon photomultipliers~\cite{Moskal:2016ztv} and new front-end electronics~\cite{marek_fee}.  

\section{Discrete symmetry tests with the J-PET detector}\label{sec:strategies}

\subsection{Measurements involving ortho-positronium spin}\label{sec:spin}
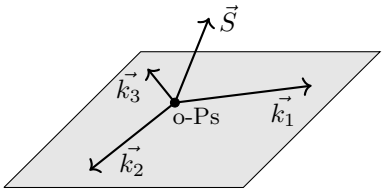
\begin{figure}[h!]
  \centering
  \begin{tikzpicture}[scale=0.45]
    \filldraw[fill=gray!20] (-5,4) -- (2,4) --  (6,8) -- (-1,8) -- (-5,4);
    \draw[fill, black] (0, 6.5)  circle [radius=0.13];
    \draw[black,thick,->] (0,6.5) node[black,yshift=-5,xshift=8]{\small o-Ps} -- (1,9) node[anchor=west] {$\vec{S}$};
    \draw[black, thick,->] (0,6.5) -- (4,7)node[midway, below,xshift=15,yshift=3] {$\vec{k_1}$};
    \draw[black, thick,->] (0,6.5) -- (-0.8,7.5)node[yshift=-7.0,xshift=-7.0] {$\vec{k_3}$};
    \draw[black, thick,->] (0,6.5) -- (-2.5,4.5)node[midway, below] {$\vec{k_2}$};
  \end{tikzpicture}
  \caption{Vectors describing the final state of a o-Ps$\to 3 \gamma$ annihilation in the o-Ps frame of reference. $\vec{S}$ denotes ortho-positronium spin and $\vec{k}_{1,2,3}$ are the momentum vectors of the annihilation photons, lying in a single plane. The operator defined in Equation~\ref{eq:op1} is a measure of angular correlation between the positronium spin and the decay plane normal vector.}\label{fig:vectors}
\end{figure}

The symmetry under reversal in time can be put to test in the o-Ps$\to 3\gamma$ decays by using the properties of T conjugation demonstrated by \mbox{Equations~\ref{eq:rel_any_operator}-\ref{eq:todd}}. Spin $\vec{S}$ of the decaying ortho-positronium atom and momenta of the three photons produced in the decay $\vec{k}_{1,2,3}$ (ordered according to their descending magnitude, i.e.\ $|\vec{k}_1| > |\vec{k}_2| > |\vec{k}_3|$) allow for construction of an angular correlation operator odd under reversal in time:
\begin{equation}
 C_{T} =  \vec{S}\cdot\left(\vec{k}_1\times \vec{k}_2\right),
  \label{eq:op1}
\end{equation}
which corresponds to an angular correlation between the positronium spin direction and the decay plane as illustrated in Figure~\ref{fig:vectors}.

Such an approach which requires estimation of the spin direction of decaying positronia was used by both previous discrete symmetry tests conducted with ortho-positronium decays~\cite{cp_positronium, cpt_positronium}. These two experiments, however, adopted different techniques to control the o-Ps spin polarization. In the $\mathcal{CP}$ violation search, positronium atoms were produced in strong external magnetic field resulting in their polarization along a thus imposed direction~\cite{cp_positronium}. A setup required to provide the magnetic field, however, was associated with a limitation of the geometrical acceptance of the detectors used. The second measurement, testing the $\mathcal{CPT}$ symmetry using the Gammasphere detector which covered almost a full solid angle, did not therefore rely on external magnetic field. Instead, positronium polarization was evaluated statistically by allowing for o-Ps atoms formation only in a single hemisphere around a point-like positron source,
resulting in estimation of the polarization along a fixed quantization axis
with an accuracy limited by
a geometrical factor of 0.5~\cite{cpt_positronium}.
Neither of the previous experiments attempted to reconstruct the position of o-Ps$\to 3\gamma$ decays, instead limiting the volume of o-Ps creation  and assuming the same origin point for all annihilations.

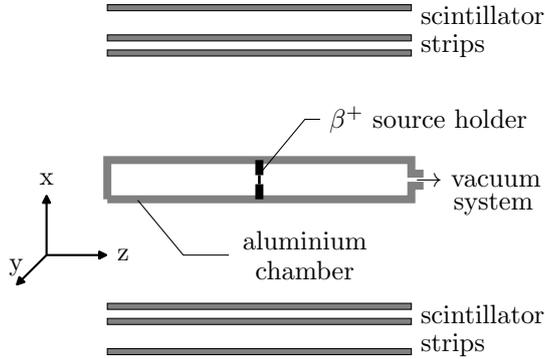
\begin{figure}[h!]
  \centering
  \begin{tikzpicture}[
  scale=0.4,
  >={Stealth[inset=0pt,length=4pt,angle'=50,round]}
  ]
  \draw[white] (-5.5,-6) rectangle (9,6);
  
  \draw[fill=black!50!white] (-5,4.1) rectangle (5,4.3);
  \draw[fill=black!50!white] (-5,4.6) rectangle (5,4.8) node[right, yshift=-4] {\normalsize strips};
  \draw[fill=black!50!white] (-5,5.6) rectangle (5,5.8) node[right, yshift=-4] {\normalsize scintillator};

  \draw[fill=black!50!white] (-5,-4.3) rectangle (5,-4.1) node[right, yshift=-4] {\normalsize scintillator};
  \draw[fill=black!50!white] (-5,-4.8) rectangle (5,-4.6) node[right, yshift=-10] {\normalsize strips};
  \draw[fill=black!50!white] (-5,-5.8) rectangle (5,-5.6);
  
  \draw[line width=3pt, black!50!white] (-5,-0.65) -- (-5,0.65) -- (5,0.65) -- (5,0.2) -- (5.4, 0.2);
  \draw[line width=3pt, black!50!white] (-5,-0.65) -- (5,-0.65) -- (5,-0.2) -- (5.4, -0.2);
  \node[] at (7.2, -0.05) {\normalsize $\rightarrow$ vacuum};
  \node[] at (7.7, -0.8) {\normalsize system};
  \node[] at (1.5, -2) {\normalsize aluminium};
  \node[] at (1.5, -3) {\normalsize chamber};
  \draw[black] (-4, -0.68) -- (-2.5, -2.5) -- (-1, -2.5);
  
  \draw[line width=3pt, black] (0, -0.65) -- (0, -0.15);
  \draw[line width=3pt, black] (0, 0.65) -- (0, 0.15);
  \draw[line width=1pt, black] (0, 0.15) -- (0, -0.15);


  \draw[black] (0.1, 0.3) -- (1.5, 2.0) -- (2.0, 2.0);
  \node[right] at (2.0, 2.0) {\normalsize $\beta^+$ source holder};


  \draw[thick, black, ->] (-7, -2.5) -- (-7,-0.5) node[above] {x};
  \draw[thick, black, ->] (-7, -2.5) -- (-5,-2.5) node[right] {z};
  \draw[thick, black, ->] (-7, -2.5) -- (-8,-3.5) node[above] {y};
  
\end{tikzpicture}
\caption{Scheme of the positronium production setup devised for positron polarization determination in J-PET experiment. Positrons are produced in a $\beta^+$ source mounted in the center of a cylindrical vacuum chamber coaxial with the detector. Positronium atoms are formed by the interaction of positrons in a porous medium covering the chamber walls. Determination of an  o-Ps$\to 3\gamma$ annihilation position in the cylinder provides an estimate of positron momentum direction.
}\label{fig:chamber}
\end{figure}

The J-PET experiment attempts to improve on the latter approach which does not require the use of external magnetic field. The statistical knowledge of spin polarization of the positrons forming o-Ps atoms can be significantly increased with a positronium production setup depicted in Figure~\ref{fig:chamber}, where polarization is estimated on an event-by-event basis instead of assuming a fixed quantization axis throughout the measurement. A trilateration-based technique of reconstructing the position of o-Ps$\to 3\gamma$ decays created for J-PET allows for estimation of the direction of positron propagation in a single event with a vector spanned by a point-like $\beta^+$~source location and the reconstructed ortho-positronium annihilation point~\cite{gajos_gps}.

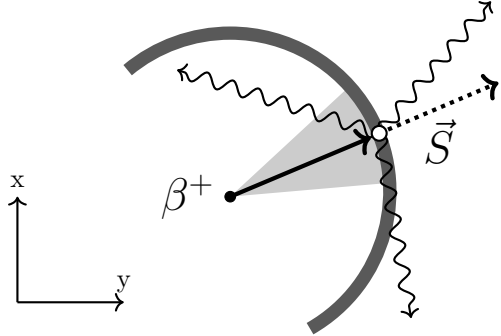
\begin{figure}[h!]
  \centering
  \begin{tikzpicture}[
  scale=1.4,  
  photon/.style={decorate, decoration={snake}, thick, ->},
  prompt_photon/.style={->, ultra thick, prompt_color},
  ->={Latex[length=16pt]}
  ]

  \coordinate (decay) at (1.4,0.6);
\coordinate (g1_a) at (-0.5, 1.2);
\coordinate (g2_a) at (2.45, 1.85);
\coordinate (g3_a) at (1.7,-1.15);

\fill[fill=gray!40!white] 
     (43:1.5) 
  -- (0,0) 
  -- (5:1.5)
  arc[start angle=5, end angle=43, radius=1.5]
  -- cycle;

\draw [darkgray!60!gray, line width=5.0, -] ($(0,0)-(-0.75,1.25)$) arc[start angle=-60, end angle=130,radius=1.5];

\draw[photon] (decay) -- (g1_a);
\draw[photon] (decay) -- (g2_a);
\draw[photon] (decay) -- (g3_a);

\draw[->, ultra thick] (0,0) -- ($(decay)-(0.08,0.04)$);
\draw[dotted, ->, ultra thick] (decay) -- ($1.8*(decay)$) node[midway, below] {\LARGE $\vec{S}$};

\draw[fill, black] (0,0) circle [radius=0.05];

\draw[thick,black,fill=white] (decay) circle [radius=0.07];

\node at (-0.4,0.0) {\LARGE $\beta^{+}$}; 

\draw[thick, black, ->] (-2, -1) -- (-2,0) node[above] {x};
\draw[thick, black, ->] (-2, -1) -- (-1,-1) node[above] {y};

\end{tikzpicture}  
  \caption{Determination of positron polarization axis using its momentum direction (black arrow) estimated using the $\beta^+$ source position and reconstructed origin of the $3\gamma$ annihilation of ortho-positronium in the chamber wall (dark gray band). The shaded region represents the angular uncertainty of positron flight direction resulting from achievable resolution of the $3\gamma$~annihilation point.}\label{fig:chamberxy}
\end{figure}

The dependence of average spin polarization of positrons (largely preserved during formation of ortho-positronium~\cite{PhysRevLett.43.1281})
on the angular accuracy of the polarization axis determination is given by $\frac{1}{2}(1+\cos\alpha)$ where $\alpha$ is the opening angle of a cone representing the uncertainty of polarization axis direction~\cite{Coleman}.
This uncertainty in J-PET results predominantly from the resolution of determination of the $3\gamma$ annihilation point as depicted in Figure~\ref{fig:chamberxy} and amounts to about \SI{15}{\degree}~\cite{gajos_gps}, resulting in a polarization decrease smaller than \SI{2}{\percent}. By contrast, in the previous measurement with Gammasphere~\cite{cpt_positronium} where the polarization axis was fixed, the same geometrical factor accounted for a \SI{50}{\percent} polarization loss.

\subsection{Measurements using polarization of photons}\label{sec:polarization}
\begin{figure}[h!]
  \centering
  \begin{tikzpicture}[scale=0.6]
    \filldraw[dashed,fill=gray!10] (1,4.5) -- (2.5,8.5) --  (8.5,8.5) -- (7,4.5) -- (1,4.5);
    \draw[black, thick, ->] (4,6.5) -- (6.7,8) node[midway, left,xshift=-3,yshift=8] {$\vec{k'_1}$};
    \draw[fill, black] (0,6.5) circle [radius=0.11] node[above, yshift=1, xshift=7]{o-Ps};
    \draw[black, thick,->] (0,6.5) -- (4,6.5)node[midway, below,xshift=5,yshift=0] {$\vec{k_1}$};
    \draw[black, thick,->] (0,6.5) -- (-0.8,7.5)node[yshift=-7.0,xshift=-7.0] {$\vec{k_3}$};
    \draw[black, thick,->] (0,6.5) -- (-2.5,4.5)node[midway, below] {$\vec{k_2}$};
    %
    \draw[black, thick, densely dotted, ->] (1.7, 6.5) -- (1.0, 8.6) node[above] {$\vec{\varepsilon}_1$};
    \draw[] (1.2, 7.9) arc(120:75:1.5);
    \node[] at (1.7, 7.6) {$\eta$};    
  \end{tikzpicture}
  \caption{Scheme of estimation of polarization vector for a photon produced in o-Ps$\to 3\gamma$ at J-PET\@. Photon of momentum $\vec{k}_1$ is scattered in one of the detection modules and a secondary interaction of the scattering product $\vec{k}'_1$ is recorded in a different scintillator strip. The most probable angle $\eta$ between the polarization vector $\vec{\varepsilon_1}$ and the scattering plane spanned by $\vec{k}_1$ and $\vec{k}'_1$ amounts to \SI{90}{\degree}.}\label{fig:vectors_polarization}
\end{figure}
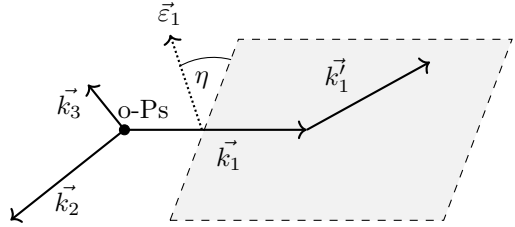

The scheme of measurement without external magnetic field for positronium polarization may be further simplified with modified choice of the measured T-odd operator.
This novel approach of testing the $\mathcal{T}$ symmetry may be pursued by J-PET with a spin-independent operator constructed for the \mbox{o-Ps$\to 3\gamma$} annihilations if the polarization vector of one of the final state photons is included~\cite{moskal_potential}:
\begin{equation}
 C'_{T} =  \vec{k}_2 \cdot \vec{\varepsilon}_1, 
\end{equation}
where $\vec{\varepsilon}_1$ denotes the electric polarization vector of the most energetic $\gamma$ quantum and $\vec{k}_2$ is the momentum of the second most energetic one. Such angular correlation operators involving photon electric polarization have never been studied in the decays of ortho-positronium. Geometry of the J-PET detector enables a measurement of $\expval{C'_{T}}$ thanks to the ability to record secondary interactions of once scattered photons from the o-Ps$\to 3\gamma$ annihilation as depicted in Figure~\ref{fig:vectors_polarization}.

\section{Test measurement with the J-PET detector}\label{sec:test}
The setup presented in Figure~\ref{fig:chamber} was constructed and fully commissioned in 2017~\cite{Czerwinski:2017ibo}. One of the first test measurements was dedicated to evaluation of the feasibility of identification and reconstruction of three-photon events. A~$^{22}$Na $\beta^+$~source was mounted inside a cylindrical vacuum chamber of \SI{14}{\centi\metre} radius. The positronium formation-enhancing medium, presently under elaboration, was not included in the measurement. Therefore, the test of $3\gamma$ event reconstruction was based on direct $3\gamma$ annihilation of positrons with electrons of the aluminium chamber walls, with a yield smaller by a factor of about 370 than the rate of o-Ps$\to 3\gamma$ annihilations expected in the final measurements with a porous medium for positronium production.
%
%

The capabilities of J-PET to select $3\gamma$ events and discriminate background arising from two-photon $e^+e^-$ annihilations as well as from accidental coincidences are based primarily on two factors:
\begin{itemize}
\item a measure of energy deposited by a photon in Compton scattering, provided by the time over threshold (TOT) values determined by the J-PET front-end electronics,
\item angular dependencies between relative azimuthal angles of recorded $\gamma$ interaction points, specific to topology of the event~\cite{kowalski_scatter_fraction}.
\end{itemize}

\begin{figure}[h!]
  \centering
  \includegraphics[width=0.45\textwidth]{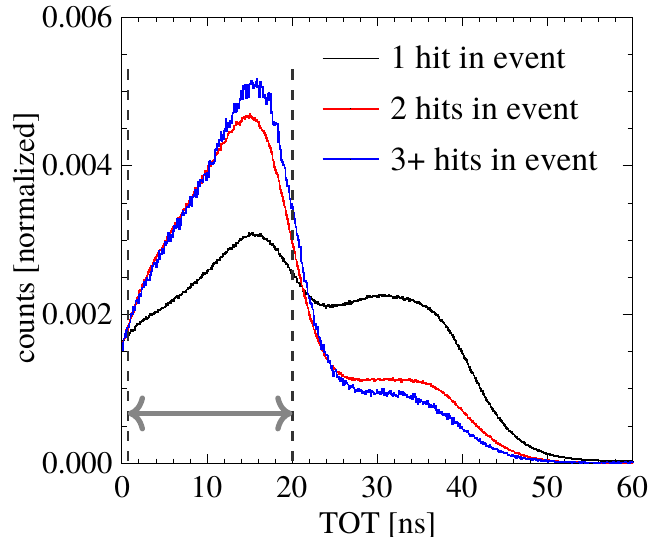}
  \caption{Distributions of time over threshold (TOT) values, for $\gamma$ interactions observed in groups of 1, 2 and more within a coincidence time window of 2.5~ns. In the samples with increasing number of coincident photons, the contribution of $\gamma$ quanta from $3\gamma$ annihilations increases with respect to other processes. Gray lines and arrow denote the region used to identify $3\gamma$ annihilation photon candidates.}\label{fig:tots}
\end{figure}

Distributions of the TOT values, after equalization of responses of each detection module, are presented in Figure~\ref{fig:tots}. Separate study of TOT distributions for $\gamma$ quanta observed in groups of 1,2 and more recorded $\gamma$ hits in scintillators within a short time window reveals the different composition of photons from 3$\gamma$ annihilations with respect to those originating from background processes such as deexcitation of the $\beta^+$ decay products from a $^{22}$Na source (1270~keV) and cosmic radiation. Two Compton edges corresponding to $511$~keV and $1270$~keV photons are clearly discernible in TOT distributions, allowing to identify candidates for interactions of 3$\gamma$ annihilation products by TOT values located below the $511$~keV Compton edge as marked with dashed lines in Figure~\ref{fig:tots}.

\begin{figure}[h!]
  \centering
  \subfloat[]{
    \includegraphics[width=0.45\textwidth]{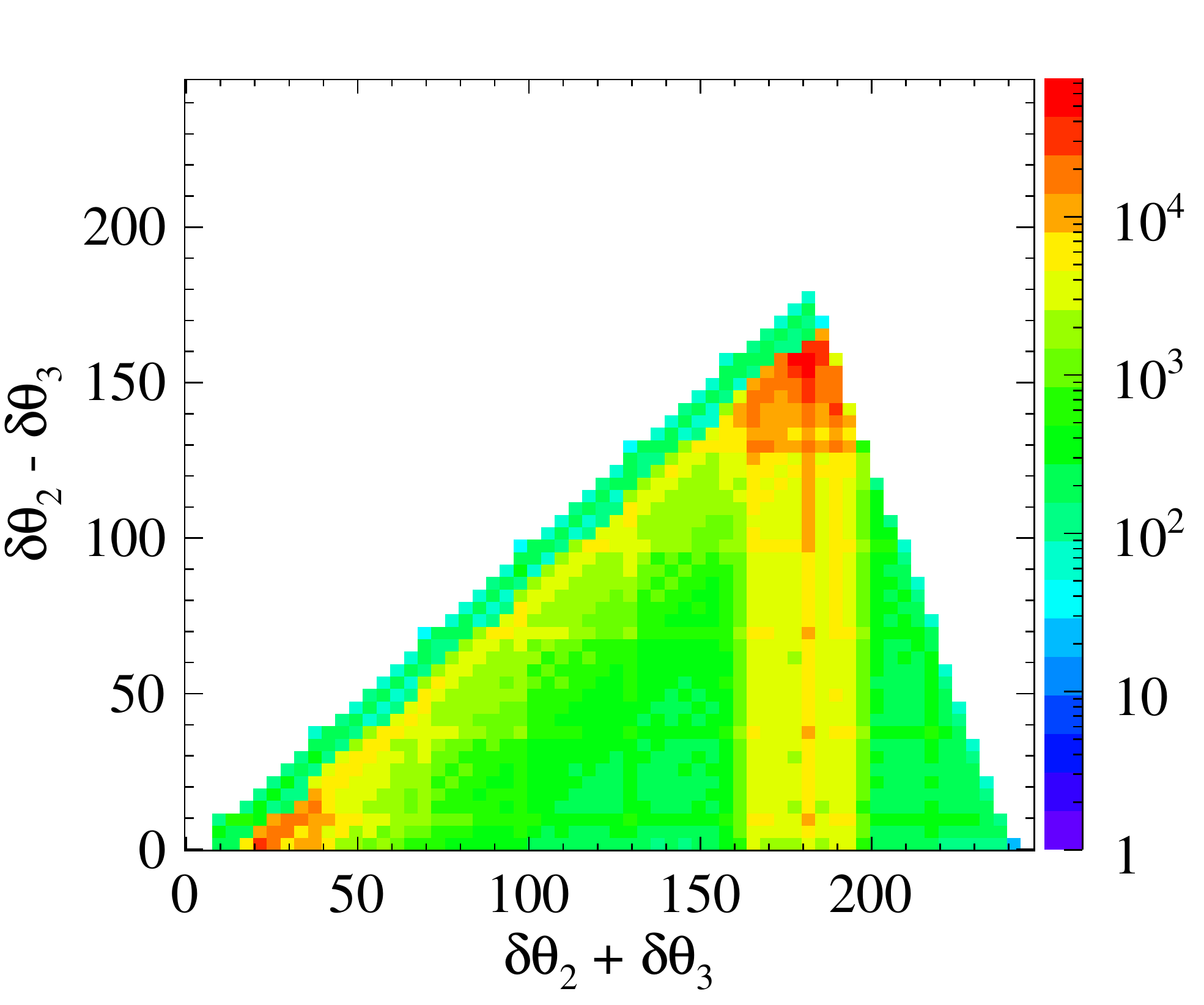}\label{fig:angles:a}
  }
  \subfloat[]{
    \includegraphics[width=0.45\textwidth]{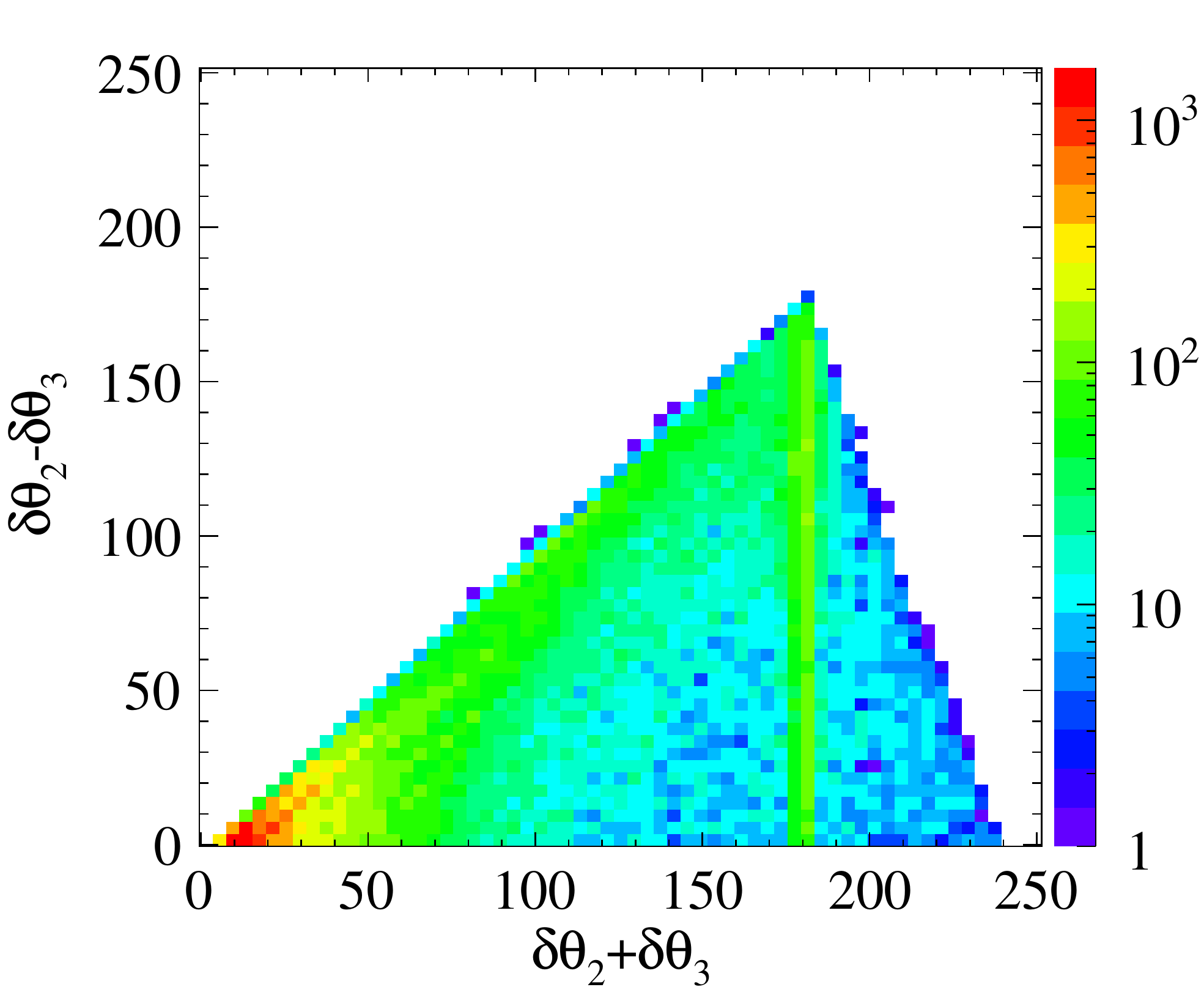}\label{fig:angles:b}
  }
  \caption{Relations between the sum and difference of two smallest relative azimuthal angles ($\delta \theta_2$ and $\delta \theta_3$ respectively) between $\gamma$ interaction points in events with three recorded interactions.
    (a) Distribution obtained in the test measurement with the aluminium chamber presented in Figure~\ref{fig:chamber}.
    For reference, the same spectrum obtained with a point-like annihilation medium located in the detector center~\cite{Czerwinski:2017ibo} is displayed in (b).
    The vertical band around $\delta\theta_2+\delta\theta_3\approx$\SI{180}{\degree} arises from two-photon annihilations and is broadened in the first case due to extensive dimensions of the annihilation chamber used.
    $3\gamma$ annihilation events are expected in the region located at the right side of the $2\gamma$ band~\cite{kowalski_scatter_fraction}.
  }\label{fig:angles}
\end{figure}

The second event selection criterion is based on the correlations between relative azimuthal angles of photon interactions recorded in the detector in cases of three interactions observed in close time coincidence. The tests performed with Monte Carlo simulations have shown that annihilations into two and three photons can be well separated using such correlations~\cite{kowalski_scatter_fraction}. An exemplary relative distribution of values constructed using these correlations, obtained with the test measurement is presented in Figure~\ref{fig:angles:a}. For a comparison, the same distribution obtained with a point-like annihilation medium used in another test measurement of J-PET is presented in Figure~\ref{fig:angles:b}.

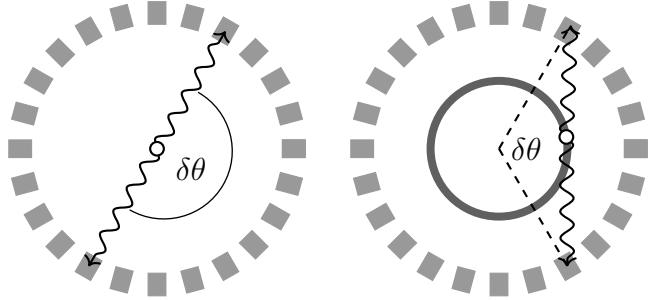
\begin{figure}[h!]
  \centering
  \begin{tikzpicture}[scale=0.45]
    \foreach \x in {0,15,...,360} \node (rect) at (\x:4.0) [draw, gray!80!white, fill, minimum width=0.1cm, minimum height=0.3cm, rotate=\x+90]{};
    \draw[decorate, decoration={snake}, thick, ->] (0,0) -- (60:4.0);
    \draw[decorate, decoration={snake}, thick, ->] (0,0) -- (240:4.0);
    \draw [thick,black,fill=white] (0,0) circle [radius=0.2];
    \draw[line width=0.5pt] (-0.8, -1.8) arc(-120:60:2.0);
    \node[] at (1.0,-0.6) {\large $\delta\theta$};
    
  \end{tikzpicture}
  \hspace{1em}
  \begin{tikzpicture}[scale=0.45]
    \foreach \x in {0,15,...,360} \node (rect) at (\x:4.0) [draw, gray!80!white, fill, minimum width=0.1cm, minimum height=0.3cm, rotate=\x+90]{};
    \draw [line width=3pt,darkgray!50!gray] (0,0) circle [radius=2.0];
    \draw[decorate, decoration={snake}, thick, ->] (10:2.0) -- (60:4.2);
    \draw[decorate, decoration={snake}, thick, ->] (10:2.0) -- (300:4.0);
    \draw[dashed, line width=0.8pt] (0,0) -- (60:4.0);
    \draw[dashed, line width=0.8pt] (0,0) -- (300:4.0);
    \draw [black,fill=white,thick] (10:2.0) circle [radius=0.2];
    \node[] at (0.8,0.0) {\large $\delta\theta$};
  \end{tikzpicture}
  \caption{Explanation of the broadening of the 2$\gamma$ annihilation band present in Figures~\ref{fig:angles:a} and~\ref{fig:angles:b} at $\delta\theta_2+\delta\theta_3\approx$\SI{180}{\degree}.
   Left: when $2\gamma$ annihilations originate in a small region in the detector center, the calculated relative azimuthal angles of detection modules which registered the photons correspond closely to actual relative angles between photons' momenta.
   Right: with $2\gamma$ annihilations taking place in the walls of an extensive-size annihilation chamber (gray band), the broadening of the band at \SI{180}{\degree} is caused by a discrepancy between the calculated and actual relative angles.
    The detector scheme and proportions are not preserved for clarity.
  }\label{fig:broadening}
\end{figure}

A sharp vertical band at $\delta\theta_2+\delta\theta_3\approx$\SI{180}{\degree} seen in Figure~\ref{fig:angles:b} originates from events corresponding to annihilations into two back-to-back photons. Broadening of this $2\gamma$ band in case of the extensive chamber is a result of the increased discrepancy between relative azimuthal angles of detection module locations used for the calculation and the actual relative angles in events originating in the walls of the cylindrical chamber as depicted schematically in Figure~\ref{fig:broadening}.

The distributions presented in Figure~\ref{fig:angles} are in good agreement with the simulation-based expectations~\cite{kowalski_scatter_fraction}. Selection of events with values of $\delta \theta_2 + \delta \theta_3$ significantly larger than \SI{180}{\degree} allows for identification of three-photon annihilations.

The aforementioned event selection techniques allowed to extract 1164 $3\gamma$~event candidates from the two-day test measurement
with a $\beta^+$ source activity of about 10 MBq placed in the center of the aluminium cylinder as depicted in Figure~\ref{fig:chamber}.
Therefore, a quantitative estimation of the achievable resolution of three photon event origin points and its impact on the positronium polarization control capabilities requires a measurement including a medium enhancing the positronium production.

\begin{figure}[h!]
  \centering
  \subfloat[]{\includegraphics[width=0.45\textwidth]{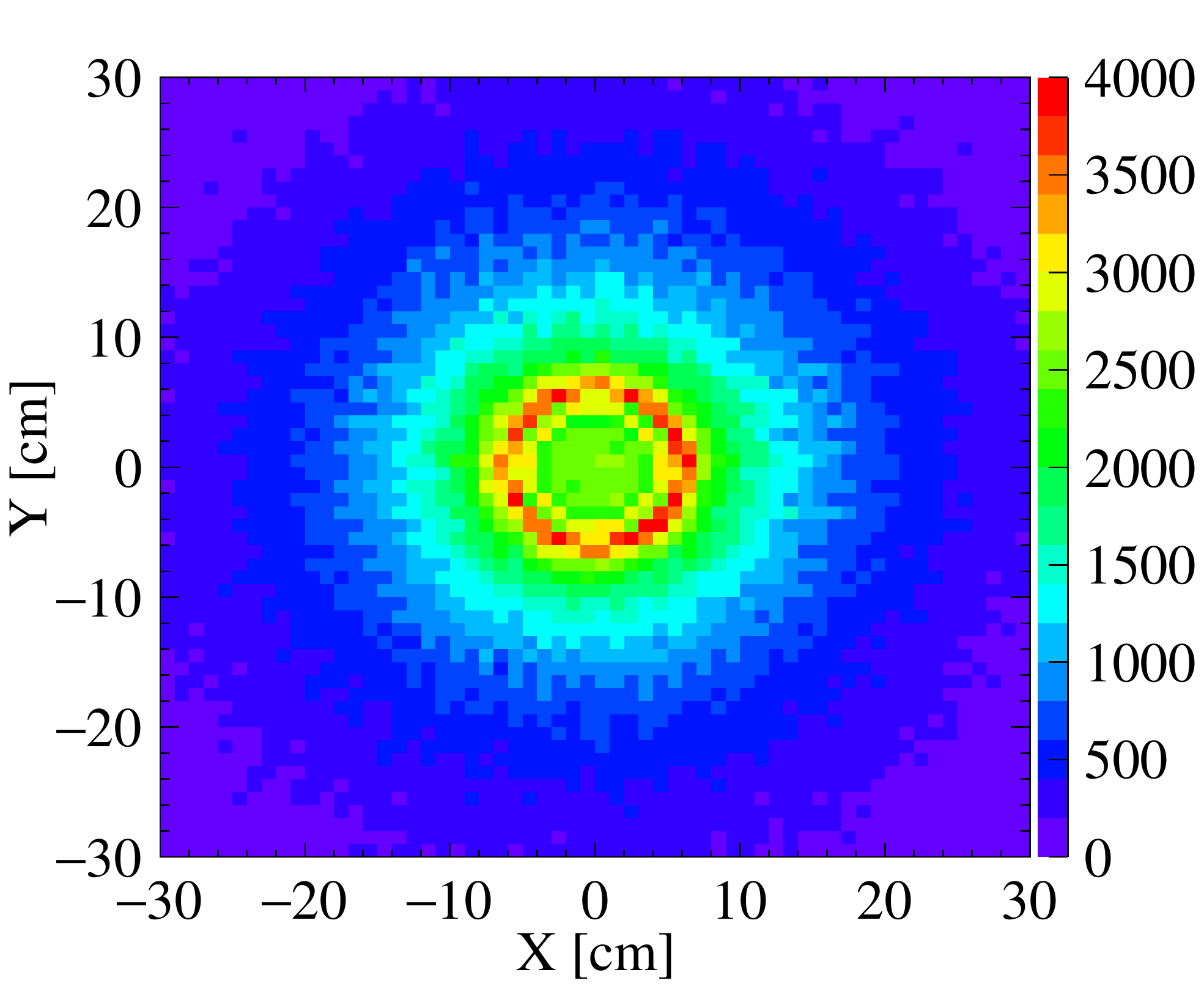}\label{Fig:2g_xy}}
  \hspace{0.05\textwidth}
  \subfloat[]{\includegraphics[width=0.45\textwidth]{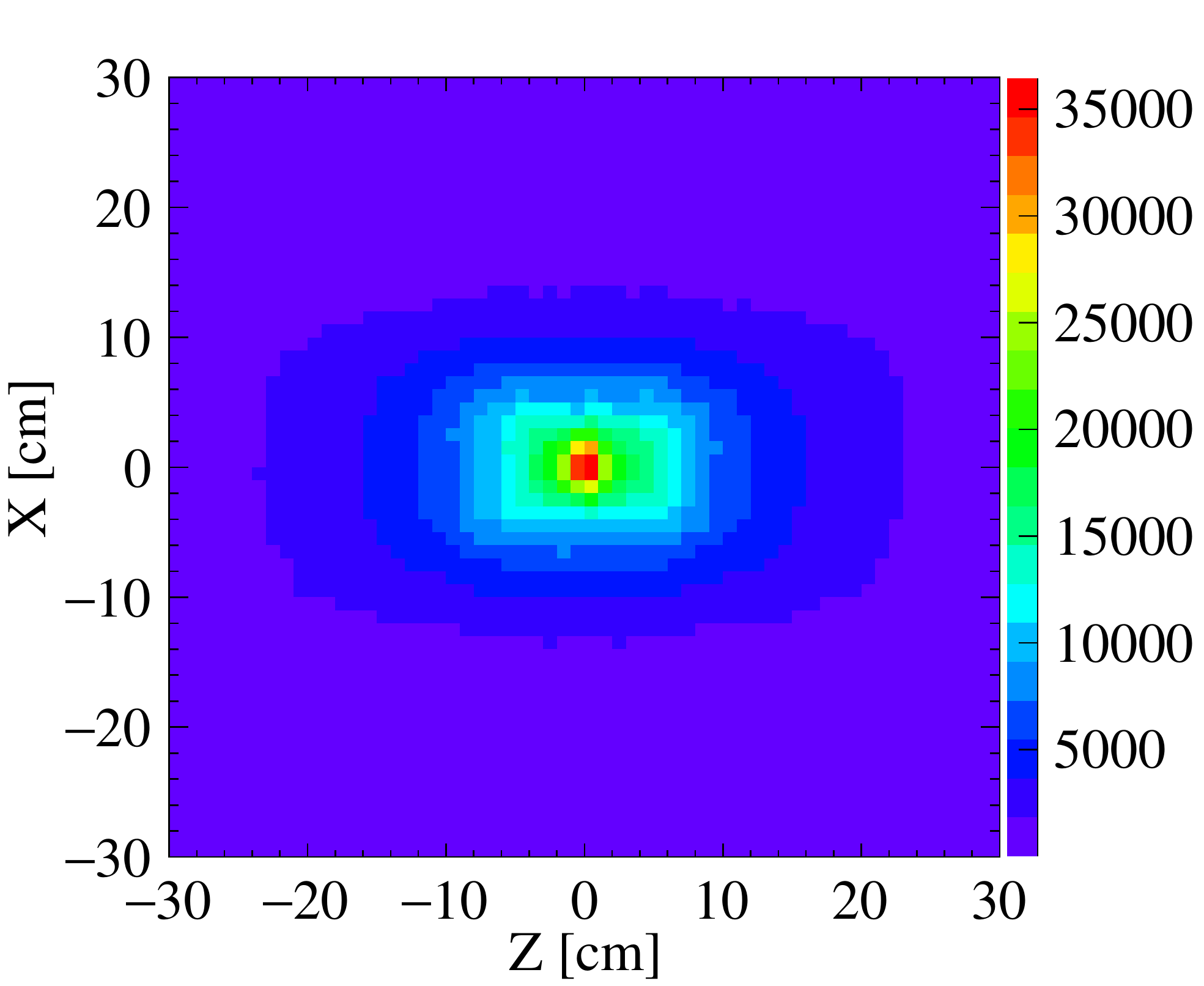}\label{Fig:2g_xz}}
  \caption{
    Tomographic images of the cylindrical chamber used in the test run of J-PET, obtained using reconstructed $e^+e^-\to 2\gamma$ annihilation events. (a) Transverse view of the chamber (the central longitudinal region of $|z|<4$~cm was excluded where the image is dominated by annihilation events originating in the setup of the $\beta^+$ source.).
    (b) Longitudinal view of the images chamber. In the central region, a strong image of the positron source and its mounting setup is visible.
  }\label{fig:2g}
\end{figure}

Resolution of the detector and its field of view was validated with a benchmark analysis of the test data performed using the abundant $2\gamma$ annihilation events. Figure~\ref{fig:2g} presents the images of the annihilation chamber obtained using 2$\gamma$ events whose selection and reconstruction was performed with the same techniques applied to medical imaging tests performed with J-PET~\cite{monika_2gamma_imaging}. Although a large part of recorded annihilations originate already in the setup holding the $\beta^+$ source, a considerable fraction of positrons reach the chamber walls. The effective longitudinal field of view of J-PET for 2$\gamma$ events which can be directly extended to $3\gamma$ annihilations due to similar geometrical constraints, spans the range of approximately $|z|<8$~cm.

\section{Summary and perspectives}\label{sec:summary}
The J-PET group attempts to perform the first search for signs of violation of the symmetry under reversal in time in the decays of positronium atoms. One of the available techniques is based on evaluation of mean values of final state observables constructed from photons' momenta and positronium spin in an o-Ps$\to 3\gamma$ annihilation with a precision enhanced with respect to the previous realization of similar measurements by determination of positronium spin distinctly for each recorded event. Moreover, the J-PET detector enables novel test by determination of a T-odd observable constructed using the momenta and polarization of photons from annihilation.

The pilot measurement conducted with the J-PET detector demonstrated the possibility to identify candidates of annihilation photons interactions in the plastic scintillator strips by means of the time over threshold measure of deposited energy and angular dependencies between relative azimuthal angles of $\gamma$ interaction points specific to event spatial topology. A preliminary selection of three-photon annihilation events yielded 1164 event candidates from a two-day test measurement with a yield reduced by a factor of about 370 within respect to the planned experiments with a porous positronium production target and a centrally-located 10~MBq source. The annihilation reconstruction resolution and  performance of the setup proposed for positron spin determination was validated with a benchmark reconstruction of two-photon annihilations. Results obtained from the test measurement confirm the feasibility of a test of symmetry under reversal in time by measurement of the angular correlation operator defined in Equation~\ref{eq:op1} without external magnetic field once a positronium production medium is used.

\section*{Conflicts of interest}
The authors declare that there is no conflict of interest regarding the publication of this paper.

\section*{Acknowledgements}
The authors acknowledge technical and administrative support of A. Heczko, M. Kajetanowicz and
W. Migda\l{}.

This work was supported by The Polish National Center for Research
and Development through grant [INNOTECH-K1/IN1/64/159174/NCBR/12], the
Foundation for Polish Science through the MPD and TEAM programmes and grants no.\
[2016/21/B/ST2/01222],\linebreak[3] [2017/25/N/NZ1/00861],
the Ministry for Science and Higher Education through grants no. [6673/IA/SP/2016],
[7150/E-338/SPUB/2017/1] and [7150/E-338/M/2017], the EU and MSHE grant no. [POIG.02.03.00-161 00013/09].
B.C.H. acknowledges support by the Austrian Science Fund ([FWF-P26783]). C.C. acknowledges a grant from the John Templeton Foundation [(ID 58158)]. The opinions expressed in this publication are those of the authors and do not necessarily reflect the views of the John Templeton Foundation.

\bibliography{refs}

\end{document}